
%
%
%
%
\input jnl
\input epsf
\def\id{\rm I}
\title{Environment--Induced Decoherence, Classicality and
Consistency of Quantum Histories}
\author{Juan Pablo Paz and Wojciech Hubert Zurek}
\affil{Theoretical Astrophysics, T6, MS-B288,
Los Alamos National Laboratory, Los Alamos, NM 87545}
\dateline
\abstract{We prove that for an open system,
in the Markovian regime, it is always possible to
construct an infinite number of non trivial sets of histories
that exactly satisfy the probability sum rules. In spite of being perfectly
consistent, these sets manifest a very non--classical behavior:
they are quite unstable under the
addition of an extra instant to the list of times defining the history.
To eliminate this feature
--whose implications for the interpretation of the formalism we
discuss--  and to achieve the stability
that characterizes the quasiclassical domain, it is necessary to
separate the instants which define the history by time intervals
significantly larger than
the typical decoherence time. In this case environment induced
superselection is very effective and the
quasiclassical domain is characterized by histories constructed with
``pointer projectors''. }

\endtitlepage

\head{1. Introduction}

{\it Environment induced superselection} \refto{Zurek82} explains why
macroscopic
objects are not observed in the majority of the quantum
states admissible in their Hilbert spaces. The basic idea
is that these objects are
impossible to isolate from their surroundings \refto{Zeh71}. The
continuous interaction with this environment
results in a process of {\it decoherence} \refto{Zurekpt}
which destroys, on a very short
{\it decoherence timescale}, the purity of nearly all of the initial
superpositions thus erasing quantum coherence between states which result in a
different evolution of their quantum environment.
Only the preferred
set of relatively stable states (or the associated sets of
observables) will exhibit
the key attribute of ``classical reality,'' which is
characterized by the
predictive power of the associated records \refto{Zurek81, EPR}.
In order to study decoherence, the analysis of the evolution of the
``reduced density matrix'' for the system (obtained from the
full density matrix by tracing out the environment variables) is often the
most convenient strategy \refto{Zurekpt, Zurekmazag}. Nevertheless the
role of the {\it records} accessible to the observers as well as the
{\it correlations} between these records and the rest of the Universe
must be recognised in the discussion of the {\it existential
interpretation} of quantum mechanics suggested by the decoherence
process \refto{Zurekmazag,Zurekptnew}.

The {\it consistent histories} formulation (proposed by Griffiths
\refto{Griffiths} and developed by
Omn\`es \refto{Omnes} and by Gell--Mann and
Hartle \refto{GMHsfi}) is based in the study of properties
of sets of quantum histories, which are represented by
time ordered sequences of projection operators. Although quantum mechanics
prevents one from
assigning probabilities to arbitrary sets of
quantum histories, it is still possible
to do so in certain cases. The validity
of the probability
sum rules --absence of quantum interference--
for sets of histories, which is one of
the properties of the classical domain, can be ascertained by analyzing
if consistency conditions are satisfied \refto{Griffiths}.
These conditions, which were originally expressed as commutation relations
between
the projectors which define historical events \refto{Griffiths, Omnes}, can
be related to the properties of an object called ``decoherence functional''
\refto{GMHsfi}.

The aim of this paper is to make contact
between these two approaches and study also some interesting
aspects of the consistent histories formulation. For this purpose, we will
apply it to a situation which is usually considered when discussing
the environment induced superselection approach. The Universe is divided into a
``system of interest'' ${\cal S}$ and the ``environment'' ${\cal E}$.
The histories of interest refer to the system
$\cal S$ only (i.e., histories will be made of projectors
${\bf P}=\id_{\cal E}\otimes P_{\cal S}$,
where $\id_{\cal E}$ is the
identity in the Hilbert space of the environment and
$P_{\cal S}$
is a projector in the Hilbert space of the system).

As the first step in our study
we will address a technical point and examine under what
conditions the decoherence functional for the above histories
can be constructed {\it entirely} from the reduced density matrix of the
system. The conclusion of our analysis is not unexpected but is still worth
mentioning: this construction
is possible only in the Markovian regime in which
the reduced density matrix satisfies a master equation which is local in
time. It is of course of interest to enquire how accurately can one assess
consistency of histories by investigating the evolution of the density
matrix when the evolution is, say, approximately Markovian. We shall not
discuss this issue here:
in the rest of the paper we will restrict to consider
situations in which the Markovian approximation is exactly valid.

As a second step, we will show how to construct
sets of histories that exactly satisfy the probability sum rules.
In particular we will analyze
histories for which the projectors are associated with the eigenstates of
the reduced density matrix. As these states form the so--called
``Schmidt basis'' \refto{Albrecht}, the above histories are going to be called
``Schmidt histories'' (although
this may be an extrapolation of the existing terminology).
We will show that, in the Markovian regime, {\it sets of Schmidt histories are
always consistent}. We will also
analyze some of their intriguing and --from the point of view of devising an
interpretation-- somewhat worrisome properties. In particular,
we will show that the events that form
Schmidt histories are themselves history--dependent: the set of projectors
defining Schmidt events in the ``next time'' $t_n$
depends on which projectors were applied in the past. These
projectors are imposed ``from the outside'' of the Universe --that is,
they do not depend just on the initial conditions or on the Hamiltonian but
primarily on arbitrary selections of, for example, the time sequence.
This implies that histories are influenced by an
``unphysical'' act of deciding what questions (which sets of projectors)
are going to be posed and when will they be posed.
This is to be contrasted with the events
which occur within the Universe and may decide, for example, existence or
non--existence of certain systems, thus making certain questions
interesting or natural.

In the original proposal of Griffiths \refto{Griffiths}, quantum histories were
defined as chains of branch independent projectors $P^{(k)}_{\alpha_k}(t_k)$
(the superindex $(k)$ labels the set of projectors and the
subindex $\alpha_k$ enumerates the different events within the complete set).
A complete set of histories was therefore characterized by a fixed choice of a
set of projectors $P^{(k)}(t_k)$ at every time. By contrast,
branch dependent histories are
chains of the form $C_\alpha=P^{(1)}_{\alpha_1}(t_1)\ldots
P^{(n,\alpha_1,\ldots,\alpha_{n-1})}_{\alpha_n}(t_n)$ where {\it the set}
of projectors used at time $t_k$ depends
on the alternatives $\alpha_1,\ldots,\alpha_{k-1}$ that
were realized in the past. Such branch dependent histories
were first considered by Omn\`es \refto{Omnes} (who called
them histories of type II). Their use
has been recently advocated by Gell--Mann and
Hartle \refto{GMHaspen}.

We will show that the ``branch dependence'' of Schmidt histories is
a consequence of an inevitable property of the reduced dynamics which we
shall call --in absence of a more concise description--
``environment induced noncommutativity'' (EINC).  Two initially
commuting density
matrices describing an open system (and corresponding, for example,
to two alternative events) will in general evolve into two non--commuting
operators as a result of the interaction with the environment.
Therefore, at some later time they will not be simultaneously diagonalizable
and will give rise to distinct sets of Schmidt states.

The existence of environment--induced noncommutativity,
which can be shown to be a very generic consequence
of the openness of the system \refto{Pazunp},
 does not seem to have been widely
appreciated (some of its
implications for consistent histories were discussed
in \refto{Zurekmazag}).
As we are going to restrict our
analysis to the Markovian regime, we will specifically address the
importance of EINC in this context. We will show that this property of
the evolution of the reduced density matrix
is in fact predicted by generic Markovian master
equations. Moreover, we will show that using such master equation one
can study the consequences of EINC in physically relevant situations.
In particular, we will consider
a system with a finite dimensional Hilbert space interacting with a large
environment. For concreteness, the system can be thought of as being formed
by a set of atomic levels and the environment as the quantized electromagnetic
field. The interaction between the system and the environment
generates spontaneous
transitions between the levels of the system. In the Markovian
approximation, the evolution
of the reduced density matrix can be modeled using the
Bloch equations \refto{louisell} which are widely used in
quantum optics \refto{cohen}. We will show that these equations predict
the existence of
environment induced non--commutativity and that this
effect can appear in a physically relevant timescale.

Finally, we will discuss some of the
consequences of environment--induced noncommutativity
for the relationship between the process of decoherence and
consistent histories approach.
We will conclude that, when EINC is strong,
Schmidt sets are not a good candidate to
describe a quasiclassical domain. This is because they are
highly unstable since the Schmidt
projectors that form a perfectly consistent set at times
$t_1, t_2, \ldots, t_n$ are generally quite
different from the ones that have to be used if the
time sequence is $t_2, \ldots, t_n$.
This rather quantum mechanical property (which could be crudely
described as ``instability under observation'')
disappears if one considers special sets of histories for which
the minimal temporal
separation $t_i-t_{i-1}$ is larger than a certain quantity.
This turns out to be crucial in achieving classicality.
Thus, Schmidt histories are defined by
a sequence of stable projectors if the differences $(t_i-t_{i-1})$
are sufficiently larger than the typical decoherence time of the system
(which is the time needed for an arbitrary initial state to become
approximately
diagonal in a fixed ``pointer'' basis).
In this way, we conclude that sets of histories associated with
``pointer states'' \refto{Zurek81} may be the basis for the
description of our quasi--classical domain. In brief, histories expressed
in terms of pointer states are stable and, when sampled on a timescale
larger than the decoherence time, they are also
approximately (although not exactly) consistent.

The paper is organized as follows:
In section 2 we briefly review both the environment induced superselection
and the consistent histories approaches. In section 3 we
analyze the
conditions under which it is possible to write the decoherence functional
in terms of the reduced density matrix of the system and
discuss the origin of environment induced non--commutativity,
In section 4 we analyze the properties of
the sets of Schmidt histories. Our conclusions are stated in
section 5.

\head{2. Environment Induced Superselection and Consistent Histories. }

Here we shall provide a brief overview of
these two approaches and of the relation between quantum and classical they
suggest. More detailed reviews of the environment--induced superselection
and the decoherence process are available elsewhere \refto{Zurekpt,
Zurekmazag}.
The paper of Griffiths \refto{Griffiths} is still an excellent introduction to
the
consistent
histories approach. In more recent publications,
Gell--Mann and Hartle \refto{GMHsfi, GMHaspen} introduce and discuss
the decoherence functional using also the sum over histories
formulation and discussing the relation with consistency. Omn\`es gives an
useful overview in his most recent review article \refto{Omnes}.

\subhead{2.1 Environment--Induced Superselection and the
Existential Interpretation.}

Quantum formalism allows for the existence of many more states of the objects
described
by it than we seem to encounter. In particular, macroscopic objects appear
to us in a small classical subset of
a much larger quantum selection available in the Hilbert space. The purpose
of environment--induced superselection is to ``outlaw'' the vast majority of
such states by appealing to their instability in presence of the environment.
The key point of this approach is simple: It starts with the realization that
there is a basic difference between
the consequences of quantum evolution for systems which are {\it closed}
(isolated
from their environments) and {\it open} (interacting with the ``rest of the
Universe''). Especially important is the fact that evolution of open
quantum systems violates the ``equal rights amendment'' guaranteed for
each and every state in the Hilbert space of a closed system by the quantum
superposition principle.
The process of decoherence affects different states differently. Some of such
states evolve essentially unperturbed by the coupling to the outside.
They form a
``preferred set of states'' in the Hilbert space of the system,
known as ``pointer basis'' which can be, in principle, found
by using the recently proposed
predictability sieve \refto{Zurekmazag,zhp}.
By contrast, superpositions of such pointer states
rapidly decay into density matrices which turn out to be --after the
characteristic decoherence time has elapsed-- given by mixtures approximately
diagonal in the pointer basis.

Thus, decoherence results in a negative selection process which dynamically
eliminates most of the superpositions. When monitored on a timescale larger
than the decoherence time, the system appears to obey an effective
environment--induced superselection
rule which prevents it from existing in the vast majority of the states.

The distinguishing feature of classical observables, the essence of ``classical
reality'', is the persistence of properties of classical systems, which can
exist in predictably evolving states and follow a trajectory which appears to
be deterministic.
Relative stability --or, more precisely, relative predictability of the
evolution of the states of open quantum systems-- emerges as a
useful criterion which can be employed to distinguish states
which can persist (or deterministically evolve into other predictably evolving
states).
This emphasis on predictable existence gives rise to an {\it existential
interpretation of quantum mechanics}. Only the states which
can {\it exist} on a timescale accessible to the observers
will be regarded by them as a part of the classical domain. Moreover, observers
are
also a part of the quantum Universe, and their perceptions are formed from
their memories --records of the past measurements. These records are states
of physical (and, generally, very open systems), which rapidly decohere and are
subject to environment--induced superselection.
Thus, the observers accessing
their own records will be restricted to perceiving their own memory in
terms of sets of preferred ``pointer states'' which can {\it exist} for a
long time. In this sense the dynamics
of the observables of the open system ``strikes twice'', on the one hand
limiting the
set of external observables to these
which have the predictability characterizing
the classical domain, while on the other hand, constraining the states of
internal records accessible to the observer.

The ultimate focus of this approach is then the persistence of correlations
between
the states of two systems --the states of memory (the records) and the
states of the measured system. Interaction with the environment is used to turn
a non--separable quantum correlation between them, represented by a state of
the form:
$$
|\Psi_{AS}>=\sum_i\gamma_i\  |A_i>~|\sigma_i>\eqno(corras)
$$
where $|A_i>$ are the record states and $|\sigma_i>$ are the
corresponding states of the system, into a classical correlation present in the
density matrix:
$$
\rho_{AS}=\sum_i |\gamma_i|^2 |A_i><A_i|\   |\sigma_i><\sigma_i|.\eqno(densmas)
$$
The environment contributes to this transition by becoming correlated with the
preferred sets of states:
$$
|\Psi_{AS}>\otimes |{\cal E}_0>\rightarrow \sum_i \gamma_i\
|A_i>~|\sigma_i>~|{\cal E}_i>,\eqno(pst)
$$
where $|{\cal E}_i>$ are orthonormal states of the environment. Tracing out of
the
environment results in a reduced density matrix $\rho_{AS}$ given by
\(densmas).

While the study of the stability of correlations is the point of departure for
the discussion of the interpretation \refto{Zurekpt,Zurekmazag},
much can be learned by studying the reduced dynamics of individual open quantum
systems such as an exactly solvable quantum harmonic oscillator
immersed in a heat bath of other oscillators. Such studies demonstrate
that the decoherence timescales are indeed very
short for macroscopic quantum objects and
that the form of the
interaction between the system and the environment has crucial influence on
the selection of the preferred set of states \refto{zhp,UZ,phz}.

The division of the Universe into subsystems is --in addition to quantum
theory--
the only crucial input. While this assumption is far from trivial, one can
argue that
it is needed to formulate the very problem of the emergence of classicality
which is being addressed by this approach \refto{Zurekmazag}. In particular,
the measurement problem cannot be stated without dividing the
Universe into an apparatus and a
measured system. Addition of an environment, if anything, makes the
discussion more realistic.

\subhead{2.2 Consistent Histories}

The basic concept used in the consistent histories approach
is obviously that of a ``history'' which is
a sequence of events defined at various moments
of time. An event in quantum mechanics can be thought of as
corresponding to a projection operator. Therefore, histories are represented
by time ordered
sequences of projection operators. In the ordinary formulation of quantum
mechanics it is always possible to assign probabilities to single
events defined by projectors, and representing (for example)
the alternative outcomes of a measurement
performed at an arbitrary time. Quantum interference prevents us from
assigning probabilities to arbitrary histories but
probabilities can be consistently
assigned to special sets of histories.
The basic idea in the consistent
histories approach is to analyze sets of histories and establish the
mathematical conditions that must be satisfied in order to be able to
define a probability measure in such set.

A history is formally represented by
a sequence of Heisenberg projectors of the form $C_\alpha=
\{P_{\alpha_1}^{(1)}(t_1), \ldots, P_{\alpha_n}^{(n)}(t_n)\}$ (where we
assume $t_1<\ldots<t_n$).
The superscript $(k)$ labels the set of projectors used
at time $t_k$ and $\alpha_k$ denotes the particular alternative. We
will consider the possibility that the set used at time $t_k$ depends
on the previous alternatives $\alpha_j, j<k$. When necessary, we will
make this dependence explicit by writing $P^{(k,\alpha_1,\ldots,
\alpha_{k-1})}_{\alpha_k}(t_k)$ but we will try to avoid using this cumbersome
notation when there is no danger of confusion.
A set of histories is said to be exhaustive if
it covers all possible alternatives at all of the different times.
Technically this is expressed by the identity
$\sum\limits_{\alpha}C_\alpha=\id$ (where $\alpha$ denotes the set of
alternatives $\alpha=\{\alpha_1,\ldots\alpha_n\}$).

In standard quantum mechanics the probability of a given event, associated
with the projector $P_i(t)$, is
computed using the formula $p(i)=Tr(P_i^\dagger(t)\rho(t_0)
P_i(t))$ (where $\rho(t_0)$ is the initial density matrix).
If we generalize this formula to histories, the natural
candidate for the
probability of the history $C_\alpha$ is
$$
p(C_\alpha)=Tr\bigl(C_\alpha^\dagger\rho(t_0)C_\alpha\bigr)\eqno(prhis)
$$
The failure of the probability sum rules can be easily seen if we
apply the formula \(prhis) to the ``coarse grained''
history $C_\beta\equiv\{C_\alpha\ {\rm or}\ C_{\alpha'}\}$. As the
logical operation ``or'' is represented by the sum of the respective
operators, we have:
$$
p(C_\alpha\ {\rm or}\ C_{\alpha'})=p(C_\alpha)+p(C_{\alpha'})+
2{\rm Re}\bigl(Tr(C_\alpha^\dagger\rho(t_0)C_{\alpha'})\bigr)\eqno(pror)
$$
The last term in \(pror) clearly violates
the probability sum rules. Thus, these
rules are satisfied in a complete set of
histories if and only if the last term vanishes for every pair of
histories in the set. To express and analyze the validity of this
condition it is convenient to define the decoherence
functional $D(\alpha,\alpha')$ as
$${
\eqalign{D(\alpha,\alpha')&\equiv Tr\bigl(C_\alpha^\dagger\rho(t_0)C_{\alpha'}
\bigr)\cr
&=  Tr\Bigl[P_{\alpha_n}^n(t_n)\ldots P_{\alpha_1}^1(t_1) \rho(t_0)
          P_{\alpha'_1}^1(t_1)\ldots
P_{\alpha'_n}^n(t_n)\Bigr]\cr}}\eqno(dfdef)
$$
The necessary and sufficient condition to define
the  probability measure in the set of
histories is now easily written as \refto{Omnes}:
$$
{\rm Re}(D(\alpha, \alpha'))=0 \quad {\rm for \quad all}\quad
\alpha\neq\alpha'\eqno(consomn)
$$
When this condition is satisfied, the set is a
{\it consistent set of histories} and the probabilities
are given in terms of the diagonal elements of the decoherence functional.
A more restrictive
condition than \(consomn)
has been proposed by Gell--Mann and Hartle \refto{GMHsfi}
which call for
the cancellation of all the non diagonal elements of the decoherence functional
and not only of its real part
$$
D(\alpha, \alpha')=0\qquad{\rm for \ all}\quad
\alpha\neq\alpha'\eqno(consgmh)
$$
(this is obviously a sufficient condition for consistency).
A brief remark about terminology is in order here.
The above consistency conditions are called ``decoherence
conditions'' by Gell--Mann and Hartle \refto{GMHsfi, GMHaspen}
who refer to \(consomn) as ``weak decoherence'' and to the
condition \(consgmh) as ``medium decoherence''. We
prefer to use the word decoherence to describe a physical
process outlined in the previous subsection. Therefore, following
Griffiths and Omn\`es \refto{Griffiths, Omnes} we will
refer to the various conditions which
assure the validity of the probability sum rules as to ``consistency
conditions''.
In practice, we
will always use the condition \(consgmh) (``medium decoherence''
in the terminology of Gell--Mann and Hartle).

Before closing this subsection let us make a remark on the definition of
the decoherence functional given in \(dfdef). In
that expression we are
obviously using the Heisenberg picture and the projectors
$P_{\alpha_k}^k(t_k)$
are defined in terms of the Schr\"odinger picture projectors as
$$
P_{\alpha_k}^k(t_k)=
U^{\dagger}(t_0,t_k)P_{\alpha_k}^k(t_0)U(t_0,t_k)\eqno(projevol)
$$
where $U(t_0,t_k)$ is the evolution operator that propagates
the state vector from $t_0$ to $t_1$. Introducing \(projevol)
into \(dfdef) we can obtain the following well known formula where
the
projectors are constant (Schr\"odinger) operators:
$$
D(\alpha, \alpha')= Tr\Bigl[P_{\alpha_n}^nU(t_{n-1},t_n)\ldots
P_{\alpha_1}^1  \rho(t_1) P_{\alpha'_1}^1
\ldots U^{\dagger}(t_{n-1},t_n)P_{\alpha'_n}^n\Bigr]\eqno(dfsh1)
$$

In the
forthcoming discussion, and for reasons that will become evident later,
it will be more convenient to use a
different expression for $D(\alpha,\alpha')$
that can be easily derived from equation \(dfsh1).
Introducing the propagator of the density matrix
(a superoperator acting in the space of operators \refto{Balescu}), which
we denote as $K_{t_i}^{t_f}$ and is defined as:
$$
K_{t_i}^{t_f}[\rho(t_i)]= U(t_i,t_f)\rho(t_i)U^{\dagger}(t_i,t_f) = \rho(t_f),
\eqno(kdef)
$$
the decoherence functional \(dfsh1) can be rewritten as:
$$
D(\alpha, \alpha')=Tr\Biggl[P_{\alpha_n}^n K_{t_{n-1}}^{t_n}\Bigl[
\ldots P_{\alpha_1}^1 K_{t_0}^{t_1}[\rho(t_0)] P_{\alpha'_1}^1 \ldots
\Bigr] P_{\alpha'_n}^n\Biggr]\eqno(dfsh2)
$$

\subhead{2.3 Decoherence, Consistency, the Quantum and the Classical.}

The purpose of this subsection is to provide a more specific motivation
for the study of the relationship
between the decoherence process and the consistency of
histories. In particular, both of them aim to
explain the emergence of the classical from the quantum substrate. The manner
in which decoherence process and the resulting environment--induced
superselection
approach this goal is quite clear. It has been briefly explained in 2.1.
When considered from the point of view of Everett's ``Many Worlds''
point of view, decoherence
defines branches. Its focus on the stable existence of the records allows one
to
understand Bohr's ``Copenhagen Interpretation'' as, in effect, an observers
memory
of
one of the Everett's branches, with the apparent collapses induced by the
effective superselection rules \refto{Zurekpt}.

The stated goals of consistent histories approach were
initially somewhat different:
consistency was invoked to discuss sequences of events in a closed evolving
quantum
system without the danger of logical contradictions \refto{Griffiths}.
However, this goal as
well as the validity of the probability sum rules are also a precondition for
the
classical behavior. Thus, at least some of the aspects of the classical domain
should be related to consistency.

However, consistency alone does not suffice to define classical
behavior \refto{GMHsfi}.
For example, given a closed system, it is always possible to find a consistent
set of
histories which are defined simply by the projectors constructed from the
evolved eigenstates of the initial density matrix. Thus, when
$$
\rho_i=\sum_i p_i |i><i|,
$$events represented by projectors
$$
\Pi_i(t_k)=|i><i|
$$
or by their direct sums can be always used to construct consistent
histories. However, when this simple algebraic algorithm is applied
to the classic test cases (such as the measurement problem or a Schr\"odinger
cat) it will result in extremely non--classical consistent histories with the
events corresponding to superpositions of various outcomes of measurements,
dead and alive cats, etc.

At the very least, consistency
would need to be supplemented by extra ingredients (which, in the
context of consistent histories interpretation, are yet
to be identified) in order to become an effective tool in studying
classicality.
Moreover, it appears likely that exact consistency may be too strong a
requirement,
and will have to be relaxed in order to be relevant for the study of
``classicality''.

On the other hand, as we will
explicitly show in the next section, the
decoherence process and the resulting environment--induced
superselection rules enforce approximate validity of the probability
sum rules for the subsystem of interest. From the perspective of the
consistent histories approach, the process of tracing over the environment
can be naturally related to a coarse grained class of histories which
correspond to sequences of projectors
${\id}_{\cal E}\otimes  P_{\alpha_k}^{(k)}(t_k)$ where
$\id_{\cal E}$ is the identity
in the Hilbert space of the environment and
$P_{\alpha_k}^{(k)}(t_k)$ acts on the Hilbert space of the system.

In what follows, we will closely investigate the connection between the
two formalisms and focus our attention on the possibility of constructing
perfectly consistent histories for the system out of the eigenstates
of the reduced density matrix. In this respect, it
is worth remembering here that the reduced density matrix can always
be instantaneously diagonalized. Its eigenstates, which are sometimes
called Schmidt states \refto{Albrecht}
are not necessarily identical (or even approximately the same) as
the pointer states: The two sets of states can be expected to coincide only
when the decoherence process has been effective which, in turn, implies
restrictions on the timescales. Thus, the time at which the Schmidt states
are calculated must be larger than the typical decoherence time scale
of the problem. In that case the Schmidt states
become independent of the details of the initial condition and
coincide with the pointer projectors.

\head{3. Consistent Histories for an Open System.}

In this section we will first establish the conditions under which
the decoherence functional can be constructed entirely in terms of the
reduced density matrix of the system. This will be shown to be possible in
the Markovian regime of the reduced evolution. In this case, we will analyze
the importance of the ``environment induced noncommutativity'' in determining
the properties of consistent histories. We should remark that our analysis of
induced noncommutativity will be restricted to
the Markovian regime despite of the fact that EINC is a
generic property of the evolution of an open system.

\subhead{3.1 Decoherence Functional for an Open System.}

When we evaluate the decoherence
functional in the histories of the system,
we have to use projectors of the form
${\id}_{\cal E}\otimes  P_{\alpha_k}^k(t_k)$ but we must remember
that the evolution between intermediate times is entirely
unitary, that $\rho_U$ is the full density matrix
and that the final
trace is over the whole Hilbert space. Thus,
the decoherence functional is obtained by tracing over the
environment at the final time while the reduced density matrix is defined
by tracing over the environment at every moment of time.
This indicates that the decoherence functional could be written in terms
of the reduced density matrix only
when taking the trace over the environment at the
end is not very different from doing it at every time. This will
be the case whenever the time evolution of the reduced density
matrix does not depend
upon the correlations that are created dynamically
between the system and the environment. The demonstration of this
simple observation can be easily done if we
decompose the final trace
$$
Tr_{\{{\cal S, E}\}}= Tr_{\cal S}\  Tr_{\cal E}
$$
and try to move the trace over the environment to the inside of the expression
for the decoherence functional. Using equation \(dfsh1) we
obtain
$$
D(\alpha, \alpha')=Tr_{\cal S}\Biggl[ P_{\alpha_n}^n Tr_{\cal E}\Bigl[O_n(t_n)
\Bigr]
P_{\alpha'_n}^n\Biggr]\eqno(firststep)
$$
where we defined $O_0(t_0)=\rho_U(t_0)$ and
$$
O_{k+1}(t_{k+1})\equiv {\it U}(t_{k},t_{k+1})
(\id_{\cal E}\otimes {\it P}_{\alpha_{k}}^{k})
{\it O}_{k}(t_k)
(\id_{\cal E}\otimes {\it P}_{\alpha'_{k}}^{k})
{\it U}^{\dagger}(t_{k},t_{k+1})
\quad 1\leq k\leq n\eqno(on)
$$
\noindent From this expression
we can now notice that the trace can be moved one more step towards the center
only if $Tr_{\cal E}(O_n(t_n))$ is
a function of $Tr_{\cal E}(O_{n-1}(t_{n-1}))$, which is the trace of an
operator
defined at $t_{n-1}$. This is not possible in
general since $Tr_{\cal E}(O_n(t_n))$ may depend on the full operator
$O_{n-1}(t_{n-1})$ and not only upon its partial traces (this is
precisely what
happens when the correlations between the system and its
environment affect the reduced dynamics of the system and
produce non Markovian effects).
We will restrict our future considerations to those cases where this is
true, or equivalently, when there is a well defined  reduced evolution
operator, denoted as $\hat K_{t_{n-1}}^{t_n}$,
acting in the following way
$$
Tr_{\cal E}(O_n(t_n))=\hat K_{t_{n-1}}^{t_n}\Bigl[ P_{\alpha_{n-1}}^{n-1}
Tr_{\cal E}\bigl(O_{n-1}(t_{n-1})\bigr)  P_{\alpha'_{n-1}}^{n-1}\Bigr]
$$
If this operator exists the decoherence functional can be written as:
$$
D(\alpha, \alpha')=Tr_{\cal S}
\Biggl[P_{\alpha_n}^n \hat K_{t_{n-1}}^{t_n}\Bigl[
\ldots P_{\alpha_1}^1 \hat K_{t_0}^{t_1}[\rho_r(t_0)]
P_{\alpha'_1}^1 \ldots
\Bigr]  P_{\alpha'_n}^n\Biggr]\eqno(dfred)
$$

This is the main equation we will use in the next section.
It is worth noting the similarity between the expression \(dfsh2)
(which is the decoherence functional for a closed system) and \(dfred), in
which all the characters are members of
the ``reduced theory''.
In the previous section
we showed that the decoherence functional for a closed system can be
written in three equivalent ways given by equations
\(dfdef), \(dfsh1) and \(dfsh2). However we can prove
that it is not possible to write the ``reduced''
decoherence functional in a way
resembling Eqs. \(dfdef) or \(dfsh1). In this sense, equation
\(dfred) is unique. In fact, only if the evolution is unitary
(as it is for the full density matrix) it is true that
to propagate the density matrix $\rho$
we just have to multiply it from left
and right with two operators
that act on the same Hilbert space
as $\rho$. This is the crucial property (see eq. \(kdef))
allowing us to show the equivalence
between equations
\(dfdef), \(dfsh1) and \(dfsh2). In the case of the reduced
density matrix this property is no longer valid since
two operators $A$
and $B$ satisfying
$$
K_{t_i}^{t_f}[\rho_r(t_i)]= A\times\rho_r(t_i)\times B = \rho_r(t_f)
\eqno(nokred)
$$
do not exist: Existence of such operators
would imply that initial pure states would remain pure forever,
which is in contradiction
with well known properties of the reduced dynamics.

Summarizing, we conclude that when the reduced evolution operator
$\hat K$ exists, the decoherence functional can be written
as \(dfred) in terms of ``reduced objects''.
Although this operator does not
exist in general,
it is also clear that there are very important cases
for which the existence of $\hat K$ is guaranteed. We will discuss those
cases in the next subsection.

\subhead{3.2 Reduced Dynamics and Environment Induced Non--Commutativity}

The existence of the reduced evolution operator $\hat K$ is a
strong requirement. For example, it implies that
the reduced density matrix satisfies a purely differential equation
(the evolution cannot have memory). Such a Markovian equation does not exist
in general since the exact master equation (the equation
for the reduced density matrix) is typically nonlocal in time.
Moreover, the existence of a local master
equation is a necessary but not a sufficient condition for the existence
of $\hat K$. In fact, there are cases for which a local but explicitly
time dependent master equation
exists and the evolution is still (weakly) non-Markovian. In those
cases the reduced
evolution operator may not exist because the correlations still play
some role in the reduced dynamics. Technically, the condition
that guarantees the existence of $\hat K$ is the {\it locality of the
Feynman--Vernon
influence functional} \refto{FV}
which enters in the path integral representation of the
decoherence functional. This implies that the master equation is
local in time and, for most realistic examples,
also has time independent coefficients. Thus, we
will restrict ourselves to consider cases for which the
influence functional is local and the master equation is time independent.
This strong assumption
will still allow us to study realistic and relevant situations.
Let us now mention two important
physical examples in
which the existence of $\hat K$ constitute a sound approximation.

The first example is the well
known linear quantum Brownian motion (QBM). In this case the system
is a particle which interacts with an environment formed by a collection
of harmonic oscillators. Assuming that the initial state does not contain
correlations between the particle and its environment, the model can be
fully
characterized by the spectral distribution
and by the initial state of the
environment (usually taken as thermal equilibrium at some
temperature $T$). In such a rather general case,
the existence of a local master equation was recently proved \refto{HuPZ1}.
It was shown that the master equation for the linear QBM is always of
second order in partial derivatives and has time dependent coefficients
that vary with temperature and with the spectral density
of the environment (time dependence in the coefficients is responsible for
all the non--Markovian effects).
A particularly important case is that of an ohmic
environment (linear spectral density) at high temperatures
($k_BT\gg\hbar\Lambda
\gg\hbar\Omega_R$, where $\Lambda$ is the high frequency cutoff of the
environment
and $\Omega_R$ is the renormalized frequency of the system). In that case,
after a short transient whose duration
is determined by $\Lambda$, the master equation for the reduced density matrix
$\rho$ reads
$$
\dot\rho=-{i\over\hbar}[H_R,\rho]-{\gamma\over{2\hbar}}
[\{p,x\},\rho]-{i\gamma\over\hbar}\bigl([x,\rho p]-[p,\rho x]\bigr)-
{{2m\gamma k_BT}\over{\hbar^2}}[x,[x,\rho]]\eqno(mastereq)
$$
Above, $H_R$ denotes the renormalized Hamiltonian of the system and
$\gamma$ is a constant that fixes the relaxation rate. Although this
equation is not valid for low temperatures,
it has been also shown \refto{UZ, phz} that in that regime
(i.e., $k_BT<\hbar\Lambda$) the high temperature approximation
remains rather accurate since the coefficients approach their
asymptotic values very fast.

A second example in which the use of a local master equation is a reasonable
approximation can be found in the domain of atomic physics and quantum
optics. In that case we consider the system to be an atom and the
environment to be formed by the infinite number of modes of the quantized
electromagnetic field. When the interaction between the system and the
environment is taken into account, a local master equation known as Bloch
equation can be derived under a number of approximations. The essential
ones are the following: absence of initial correlations
between the system and the environment, Markovian behavior (very short
life time of correlations in the environment), weak coupling (the equations
are valid to second order in an expansion in the coupling constant)
and rotating wave approximation (by which rapidly varying terms ---counter
rotating--- are supposed to average to zero). If we denote with $|n>$ the
eigenstates of $H_0$, the hamiltonian of the isolated atom, Bloch equations
(in the interaction picture associated with $H_0$) read:
$$
\dot\rho_{nm}=-{i\over\hbar}[H_{d},\rho]_{nm}+
\delta_{nm}\sum_kw_{nk}\rho_{kk}-\Gamma_{nm}\rho_{nm}\eqno(bloch)
$$
Here, the driving hamiltonian $H_d$
accounts for the coherent effects associated
with the interaction between the atom and the electromagnetic field (such as
coherent driving producing Rabi oscillations). The constants $w_{nk}$ are
transition
rates that, in the absence of driving, determine the evolution of the
diagonal elements of $\rho$ (populations)
while $\Gamma_{nm}$ are related to decay rates
that affect the evolution of
the nondiagonal elements (coherences).
These constants can be, in principle, expressed
in terms of some microscopic model and cannot be thought of as being
independent of each other because of the fluctuation--dissipation relations
(and the conservation of probability which implies that
$\Gamma_{nn}=\sum_kw_{kn}$).

Let us now discuss
one of the most remarkable features of the reduced dynamic associated with
the above master equations: the existence of
``environment induced non--commutativity'' (EINC): The existence of EINC is a
consequence of the non--unitarity of the reduced dynamics which
does not necessarily
preserve the commutation relations. Two initial states
that satisfy
$$
[\rho_a(0),\rho_b(0)]=0
$$
may evolve in such a way that
$$
[\rho_a(t),\rho_b(t)]\neq 0
$$
This is EINC, a property with
important consequences in determining the qualitative
nature of some interesting sets of consistent histories that we will
consider in the next section. It is worth noting that this effect
takes place on a rather special timescale.
Commutativity -- rather than non--commutativity --
is induced on a timescale which is very much shorter than the time needed
to approach equilibrium. This is the decoherence timescale.
Thus, on that timescale
every initial state will become approximately diagonal in the same pointer
basis
(and therefore the final states will always commute). Therefore, the timescale
on which EINC is most important is shorter than the decoherence timescale.

The existence of EINC is a prediction of both master equations
\(mastereq) and \(bloch). In particular, we can  show that in
the linear QBM the commutator changes as
$$
{{d\  }\over{dt}}[\rho_a(t),\rho_b(t)]\propto {{2m\gamma k_BT}\over{\hbar^2}}
[[\rho_a(t),x],[\rho_b(t),x]].\eqno(eincqbm)
$$
Similarly, it is simple to show that Bloch equations also predict the existence
of EINC with the only restriction that the dimension of the system's
Hilbert space be greater than two (no EINC for a spin $1/2$ system).
Finally, we should stress that in
order to be really sure about the physical nature of the EINC predicted from
the master equations \(mastereq) and \(bloch), we still need
to show that the effect occurs on a timescale which is compatible with
the ones used to obtain those equations.
We will illustrate that this is indeed the case using a specific example
in the next section.

\head{4. Consistency, Decoherence and Classicality}

\subhead{4.1 Schmidt Histories Are Consistent}

Using equation \(dfred) for the reduced decoherence
functional, it is very simple to find a systematic way of constructing an
infinite number of consistent sets of histories. The method
gives a clear prescription for chosing the sets of
projectors $P^k_{\alpha_k}$ that guarantee perfect consistency.
Given a time sequence $t_1, t_2,\ldots,t_n$, the decoherence
functional for the histories of the system
$$
D(\alpha, \alpha')=Tr_{\cal S}
\Biggl[ P_{\alpha_n}^n \hat K_{t_{n-1}}^{t_n}\Bigl[
\ldots  P_{\alpha_1}^1 \hat K_{t_0}^{t_1}[\rho_r(t_0)]
P_{\alpha'_1}^1 \ldots
\Bigr] P_{\alpha'_n}^n\Biggr]\eqno(dfred2)
$$
is automatically diagonal in its first index if we choose
the projectors $P^1_{\alpha_1}$ in such a way that they commute with the
reduced
density matrix at time $t_1$. These projectors (if chosen to be one
dimensional) are associated with the instantaneous
eigenstates of the reduced density matrix (the so--called Schmidt basis).
Doing so, the decoherence functional reads
$$
D(\alpha, \alpha')=\delta_{\alpha_1,\alpha'_1}
Tr_{\cal S}\Biggl[ P_{\alpha_n}^n \hat K_{t_{n-1}}^{t_n}\Bigl[
\ldots  P_{\alpha_2}^2\hat K_{t_1}^{t_2}[
P_{\alpha_1}^1 \rho_r(t_1) P_{\alpha_1}^1]  P_{\alpha'_2}^2 \ldots
\Bigr] P_{\alpha'_n}^n\Biggr]\eqno(dfred3)
$$
Analogously, to achieve diagonality in the second
index of the decoherence functional, we should choose the
projectors $P^2_{\alpha_2}$ in such a way that they commute with the
path projected reduced density matrix
$\hat K_{t_1}^{t_2}[ P_{\alpha_1}^1 \rho_r(t_1) P_{\alpha_1}^1]$.
However, due to the existence of environment induced non--commutativity,
the eigenstates of the path projected density matrix will generally
depend on the alternative $\alpha_1$. Therefore, the set of
projectors chosen at time $t_2$ will generally
depend on the previous alternatives and the history will be branch dependent.

This procedure can be implemented recursively for arbitrarily many steps. It
will produce a set of branch dependent histories
for which the decoherence functional is automatically diagonal.
At time $t_k$, the projectors are
associated with the eigenvectors of the path projected reduced density matrix
$\hat K_{t_{k-1}}^{t_{k}}\bigl[\ldots\hat K_{t_1}^{t_2}[P_{\alpha_1}^1
\rho_r(t_1) P_{\alpha_1}^1] \ldots\bigr]$. These projectors
always exist and in general, due to EINC, depend
upon the alternatives $\alpha_1, \ldots, \alpha_{k-1}$. As we mentioned
above, we will refer to them as ``Schmidt histories''.
It is important to realize that by
following the above procedure we
can construct an infinite number of different sets of histories
all of which are
exactly consistent. Thus, we can obtain a different set just
by choosing a different sequence of historical instants $\{t_k\}$.
Moreover, by changing the time sequence we may
drastically
change the sets of projectors. In this sense, these sets are highly
unstable.

To illustrate this point and clarify the nature of the instability let us
imagine that we follow the above procedure and
construct a consistent set of histories
specifying projectors at
times $t_1, t_2, \ldots, t_n$. The histories belonging to this set are
strings of the form
$P^{(1)}_{\alpha_1}(t_1) P^{(\alpha_1,2)}_{\alpha_2}(t_2)\ldots
P^{(n,\alpha_1,\ldots,\alpha_{n-1})}_{\alpha_n}(t_n)$.
Let us now construct another consistent set by using the same method
but specifying histories at times $t_2, \ldots, t_n$.
In this way we obtain histories which are strings of the
form $\tilde P^{(2)}_{\alpha_2}(t_2)\ldots
\tilde P^{(n,\alpha_2,\ldots,\alpha_{n-1})}_{\alpha_n}(t_n)$. The sets are
unstable because,
due to the ``environment induced non--commutativity'', the projectors
$P^k_{\alpha_k}$ are different from
the projectors $\tilde P^k_{\alpha_k}$.
In fact, in the second case the set of
projectors we must use at $t_2$ depends only upon the initial
density matrix while in the first case may strongly depend on the
alternatives $\alpha_1$.
In the next subsection we will illustrate this fact with an example
that demonstrates that the effect is real and can be rather large.

The natural question to
ask is if there is some situation in which the above
diagonalization procedure generates a unique (and stable) output. This is
going to be the case only when the eigenbasis of the ``path
projected reduced density matrix '' at time $t_k$ (i.e.,
$\hat K_{t_{k-1}}^{t_{k}}\bigl[\ldots\hat K_{t_1}^{t_2}[P_{\alpha_1}^1
\rho_r(t_1) P_{\alpha_1}^1] \ldots\bigr]$) is independent
of the path projected reduced density matrix at time $t_{k-1}$.
This requirement is satisfied when there exists a stable pointer
basis \refto{Zurekmazag}: We need the
environment to help
select a preferred (and stable) set of states. However,
this can only happen if we wait long enough between the intermediate
times for which we specify the history. Roughly speaking, the
difference $\Delta t=t_i-t_{i-1}$ must be larger than the typical
decoherence timescale $\tau_{dec}$
of the problem.

\subhead{4.2 The Importance Of Environment Induced Non--Comutativity: An
Example}

We will analyze here a particular example that illustrates the importance of
EINC in producing consistent histories which may be highly unstable. For
simplicity, we will use Bloch equations and consider a system with
a low dimensional Hilbert space. As we have mentioned above,
Bloch equations cannot result in
EINC if the dimension of the system's Hilbert space
is equal to two. Thus, we need at least three dimensions. However, we are
also interested in showing an example in which EINC takes place in a timescale
for which Bloch equations are valid. This implies, roughly speaking, that
the interesting effect should take place in a timescale
longer than the lifetimes $\Gamma_{nn}^{-1}$.
It is simple to show that this cannot be done in a three dimensional example.
Thus, we will take our system to have a Hilbert space with four dimensions.
We will consider the simplest situation
in which all the levels are stable except one (say $|4>$) which can decay only
to the ground state (say $|1>$). Thus, in this case we see --neglecting
induced emission and absorption processes-- that all the coefficients entering
in the Bloch equation are either identically zero or given by:
$$
\Gamma_{44}=2\Gamma_{14}=2\Gamma_{24}=2\Gamma_{34}=w_{14}\equiv\Gamma.
\eqno(blochcoef)
$$
It is interesting to note that the decoherence timescale
of this system can be controlled in a very simple way. In the absence of
external driving (i.e., $H_d=0$) the nondiagonal elements $\rho_{k4}$ will
disappear in a timescale related to $\Gamma^{-1}$. However, as there is only
one dissipative channel, the
density matrix will not become diagonal in the three dimensional subspace
generated by $\{|k>,k=1,2,3\}$. This situation can be changed if one
introduces
a coherent driving  by coupling the system to intense laser fields which are
in resonance with the transitions between state $|4>$ and other states.
The intensities of the different laser fields control the frequencies of the
Rabi oscillations and these frequencies control the decoherence timescale
of the system. In our example we use this idea to make the decoherence
timescale rather large (this allows us to observe EINC on a reasonable
timescale). In particular, we consider the following simple driving terms:
$$
H_d=\Omega_{14} |1><4| + \Omega_{13} |1><3| +\Omega_{12}|1><2|
+{\rm h.c.}\eqno(hdriv)
$$
Using this hamiltonian in
Bloch equations (which were integrated numerically),
we demonstrated the existence of EINC for a rather robust
set of parameters and for relevant timescales. An
example is displayed in figure 1 in which
approximately $40\%$ of the branches show significant degree of instability.
We remark that the parameters we used are rather reasonable from the point
of view of the systems for which Bloch equations are typically used
in atomic physics.
\vfill\eject

\hskip 1.2truein
\epsfysize=3.5truein
\epsfbox{tree3.eps}

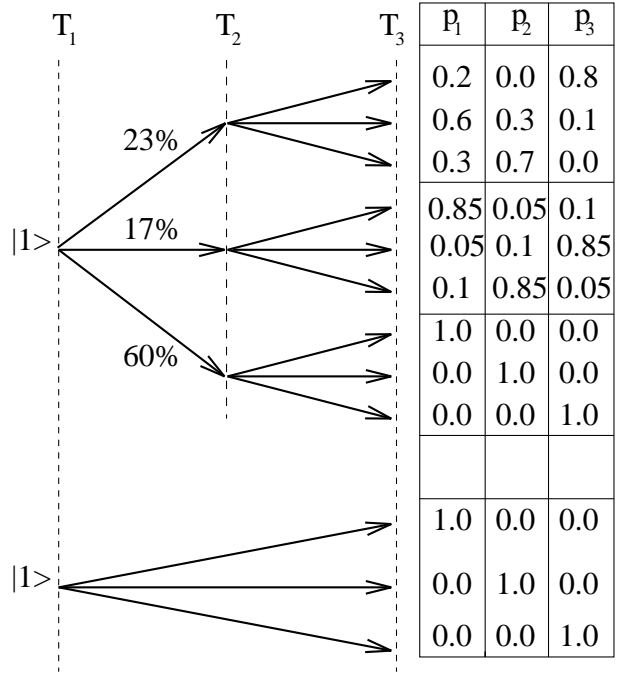
{\narrower\narrower\singlespace\noindent {\bf Figure 1:} Consistent Schmidt
histories for a system
with a four dimensional Hilbert space
are represented by a branching diagram.
This example corresponds to a system described by Bloch equations \(bloch).
The coefficients defining the driving Hamiltonian
\(hdriv) are
$\Omega_{12}=5,\ \Omega_{13}=10,\ \Omega_{14}=50$ and the
decay rate \(blochcoef) is $\Gamma=3000$ (measured in
units in which the time separation $T_2-T_1=T_3-T_2$ is set to unity).
The origin of the diagram corresponds to the state $|1>$, which is
the state of the system at time $T_1$ (we do not
draw the fourth branch corresponding to the unstable state $|4>$ since it has
zero
probability in our example).  The tabulated $p_i$'s
are the projection of the state defining each branch
onto the basis $|i>$, {\it i.e.} $p_i=|<i|\Psi>|^2$.
When the histories are constructed
at times $T_1, T_3$, there are three consistent branches corresponding to
the states $|i>, \ i=1,2,3$ ($T_3-T_1$ is larger than the decoherence
time and the states $|i>$ form a pointer basis).
On the contrary, when the histories are
constructed at times $T_1,T_2,T_3$, there are nine consistent
branches. In the first six ones (which correspond to $40\%$ of the total) the
effect of environment induced noncommutativity is important: the states
associated with the different consistent branches form a basis
of Hilbert space which is different from the one formed by the $|i>$ vectors.
The
quantum instability of the Schmidt histories is easily
noticeable.  \smallskip}

\head{5. Conclusions}

Let us summarize our results.
We analyzed first the conditions under which the decoherence functional
can be written entirely in terms of ``reduced'' quantities and showed
that this can be done when the correlations dynamically
created between the system and the environment do not affect the future
evolution of the reduced density matrix. As can be explicitly shown,
this is the case for the high temperature limit of the Caldeira--Leggett
model of ohmic dissipation and in any other case for which the Feynman--Vernon
influence functional is local in time.
Expressing the decoherence functional in terms of elements of the reduced
dynamics and using the properties of the reduced evolution operator
we proved that it is always possible to construct an infinite number
of sets of perfectly consistent histories. We also showed
that these sets are generally rather unstable under ``observations''
since by deleting one of the times at which the histories are defined,
the projection operators defining
the consistent histories may be substantially changed.

In discussing the classical limit using the
consistent histories approach various
concepts are usually brought together. The first one is consistency, which in
this formulation is the primary criterion. Coarse graining is
usually invoked as a necessary way of achieving consistency. This is
even the case when a separation of relevant and irrelevant degrees of freedom
is
made. In fact, in discussions of Caldeira--Leggett type of models it is usually
argued that to achieve consistency for histories of the Brownian particle one
needs to introduce some degree of spatial coarse graining \refto{DowkerHall}
and consider histories defined by a sequence of ``slits'' characterized
by some widths. In this case, the width of the slits is associated with the
degree of coarse graining.
However, as we explicitly showed here, such coarse grainings
are {\it not necessary} to achieve consistency. In fact, the Schmidt
projectors can be one dimensional and still define consistent histories. The
problem with the usual argument in favor of coarse graining as a way to
achieve consistency is that it uses an essential extra ingredient that
remains {\it hidden} for the most part. Thus, a
spatial coarse graining is needed {\it only} if one restricts to consider
very special histories constructed with special
classes of projection operators:
position projections for example. But there
is nothing in the consistent histories approach telling us that we must
like a set of projectors better than another! The extra ingredient that makes
us think that is ``natural'' to describe the world around us using position
projectors (or any other set of projectors we happen to like)  has nothing
to do with consistency. This extra ingredient is the crucial one in defining
the
quasi--classical domain.

As we discussed in section 2.3,
there are very simple ways of achieving consistency for a closed system.
The sets one constructs in this way are based on the use of projection
operators that are blatantly non-classical.
Our results show that for an open system (which interacts with an external
environment) the situation is much the same. Consistency is achieved
easily by means of the Schmidt histories. However, these histories have
no relation with the one describing a sensible quasi-classical domain.

The Schmidt histories we discussed provide an interesting example
that may help us to disentangle the many concepts that enter in the definition
of
a quasiclassical domain. On the one hand they are consistent but do not
require any (spatial) coarse graining.
On the other hand, although we cannot
prove it rigorously, it is likely that this set will also satisfy other
criteria that have been advanced so far (and in a less rigorous manner)
to characterize the quasi--classical domain
\refto{GMHsfi, GMHsingap}. In fact, the set of Schmidt
histories is likely to be ``full'' since
to every history of the system there should exists a projection operator
in the complete Hilbert space.
Schmidt histories are therefore an example of a set which
is consistent, rather fine grained (for the system) and most likely
full. Despite all these properties
the set is still very non--classical ! This is, of course, unless
we require {\it predictability} (or stability of the set under addition of
extra intermediate times) in which case we need to require the
separation between time slices to be
larger than the typical decoherence time. In that case,
the set becomes stable and
the projectors are determined (by the environment and not by us!) to be
the ones associated with the pointer basis.

A further conclusion one can draw from our paper is the following:
In working within the framework of the consistent
histories approach one may be tempted to think that by looking for
sets of histories that satisfy the consistency conditions (or other
stronger versions like those invoked by Gell--Mann and Hartle) one is
trying to find ``real'' histories that are ``out there'' in some
vague way. In saying this, we do not claim that the original proponents
of the formalism really made this assertion: we are just discussing what we
believe is part of the informal folklore of the field. In some sense,
consistent histories would be the ``natural'' way of looking at the
system: finding the right projectors that, when used to describe the
system, don't ``damage it'' in any way. However, our example proves that
this is certainly not the case. Consistent histories are not ``real'' in
any sense. The projectors used in constructing them must be chosen
from the outside, by us the physicists, and have a decisive
effect on the consistent histories of the quantum system.
In our example, we could decide to add an extra instant to the list of
times defining the stroboscopic history and by this simple act we would
have to change completely the description of the system!
Reality is a subtle concept that has been debated over the years in
many physics texts and does not have a clear definition. In Einstein's view,
an essential ingredient characterizing it is {\it predictability}. In
that sense, histories could be considered to be real if, apart from
the consistency condition, they are predictable behaving in
a stable way. As we showed, this is the case if histories are
constructed with pointer states.

Consistent histories interpretation was introduced by Griffiths
\refto{Griffiths}
and pursued by others to allow for a discussion of the quantum evolutions
without reference to ``measurements'' or ``collapses'' of the
wave function. The difficulties we have pointed out in our discussion appear to
stem not from attempting to achieve this goal, but from retaining the
key elements of the formal machinery of ``measurements'' (such as projection
operators acting at well defined instants of time) which then tend to influence
histories in a distinctly non--classical manner. Instead of
introducing such formal constructs ``from the outside'' of the
investigated quantum Universe, one might search for the equivalents ``on the
inside'', in the structure of the correlations between the quantum systems.
Instead of projection operators, one would then have ``records''
--relatively stable states, which, owing to the nature of the quantum
dynamics, retain their correlations with the observables of other
quantum systems. This program is, of course, embodied in the environment
induced superselection approach.

\subhead{Acknowledgements}

Part of this work was done while we were participating in
the workshop on Decoherence and the Physics of Information held at the Aspen
Center for Physics during June of1992. We would like to thank the
Center for the hospitality and all
the participants of the workshop for all their
many comments. In particular we would like to thank Murray Gell--Mann
and Jim Hartle for many interesting discussions.

\references

\refis{EPR} Predictability is indeed a key in the notion of
physical reality used in the classical EPR paper,
A. Einstein, B. Podolsky and N. Rosen,
Phys Rev {\bf 47}, 777 (1935). There,  the ``elements of
reality'' are defined as follows:
{\it If, without in any way disturbing a system, we can predict
with certainty the value of a physical quantity, then there exists an element
of physical reality corresponding to this physical quantity}.

\refis{UZ} W. G. Unruh and W. H. Zurek, Phys. Rev. {\bf D 40}, 1071 (1989).

\refis{GMHsfi}M. Gell-Mann and J. B. Hartle, in {\it Complexity, Entropy
and the
Physics of Information}, ed. W. Zurek, Vol. IX (Addison-Wesley, Reading, 1990).

\refis{GMHaspen} M. Gell--Mann and J. B. Hartle, ``Classical Equations
for Quantum Systems'', Phys Rev {\bf D}, (1993) to appear.

\refis{GMHsingap} M. Gell--Mann and J. B. Hartle, In {\it Proceedings of the
25th international conference on high energy physics}, Singapore 1991. World
Scientific.

\refis{Zurekptnew} W. H. Zurek, Physics Today (1993), to appear.

\refis{Zurek81} W.H. Zurek, Phys. Rev {\bf D 24}, 1516 (1981).

\refis{Zurek82} W.H. Zurek, Phys. Rev {\bf D 26}, 1862 (1982).

\refis{Zurek82procc} W.H. Zurek in
{\it Quantum Optics, Experimental Gravitation and Measurement
Theory},
Proceedings of the  NATO Advanced Study Institute, edited by P. Meystre and
M.O Scully (NATO ASI Series B, Vol. 94) (Plenum, New York, 1982).

\refis{Zurekpt} W.H. Zurek, Physics Today {\bf 44}, 36 (1991).

\refis{Zurekmazag} W. H. Zurek, Prog. Theor. Phys. (1993) to appear;
``Preferred Sets of States, Predictability,
Classicality, and the Environment Induced Decoherence'',
to appear in {\it ``Physical Origins of Time Asymmetry''}, J. Halliwell et al
ed., Cambridge Univ. Press (1993), in press.

\refis{Griffiths} R. Griffiths, J. Stat. Phys. {\bf36} , 219 (1984).

\refis{Omnes} R. Omn\`es, J.Stat. Phys. {\bf53}, 893 (1988)
; {\it ibid} {\bf 53}, 933 (1988); {\it ibid.}{\bf53}, 957(1988);Ann.
Phys. {\bf201}, 354 (1990); Rev. Mod. Phys. {\bf 64}, 339 (1992).

\refis{zhp} W.H. Zurek, S. Habib and J.P. Paz, Physical Review Letters
{\bf 70}, 1187 (1993).

\refis{Zeh71} H.D. Zeh, Found. Phys {\bf 1}, 69 (1971); E. Joos and
H. D. Zeh, Z. Phys. {\bf B 59}, 223 (1985).

\refis{DowkerHall} F. Dowker and J.J. Halliwell,
Phys. Rev. {\bf D 46}, 1580 (1992).

\refis{phz} J. P. Paz, S. Habib and W. H. Zurek, Phys Rev {\bf D 47}, 488
(1993).

\refis{Pazunp} J. P. Paz and W. Zurek, (1992) unpublished.

\refis{HuPZ1} B.L. Hu, J. P. Paz and Y. Zhang, Phys. Rev. {\bf D 45},
2843, (1992); {\it ibid} {\bf D 47}, 1576 (1993).

\refis{Balescu} R. Balescu, {\it ``Equilibrium and Nonequilibrium
Statistical Mechanics''} (Wiley, New York, 1975).

\refis{louisell} W.H. Louisell, {\it ``Quantum statistical properties  of
radiation''}, (J.Wiley \& Sons, New York, 1973).

\refis{cohen} C. Cohen Tannoudji et al, {\it ``Atom Photon Interactions''}
(J.Wiley \& Sons, New York, 1992).

\refis{CLphysica}A. O. Caldeira and A. J. Leggett, Physica 121A, 587 (1983)

\refis{FV} R. P. Feynman and F. L. Vernon, Ann. Phys. {\bf 24}, 118 (1963).

\refis{Albrecht} A. Albrecht, Phys. Rev. {\bf D 46}, 5504 (1992);
``Following a collapsing wave function'', Phys. Rev {\bf D}, (1993) to
appear.

\endreferences
\end